\documentstyle[prl,aps,twocolumn]{revtex}

\begin{document}
\title{Transition Temperature of a Uniform Imperfect Bose Gas}
\author{Kerson Huang\thanks{This work was supported in part by a DOE cooperative agreement
DE-FC02-94ER40818.\hfil MIT-CTP~2851}}
\address{Center for Theoretical Physics and Physics Department\\
Massachusetts Institute of Technology, Cambridge, MA 02139}
\date{\today}
\maketitle

\begin{abstract}\noindent 
We calculate the transition temperature of a uniform dilute Bose
gas with repulsive interactions, using a known virial expansion of the
equation of state. We find that the transition temperature is higher than
that of an ideal gas, with a fractional increase $K_0 (na^3)^{1/6}$,
where $n$ is the density and $a$ is the S-wave scattering length, and
$K_0$ is a constant given in the paper. This disagrees with all existing
results, analytical or numerical. It agrees exactly in magnitude with a
result due to Toyoda, but has the opposite sign.
\end{abstract}


The weakly interacting Bose gas is an old subject that has found new life
since the experimental discovery of Bose-Einstein condensation in ultra-cold
trapped atoms. Although there is a vast literature on the subject, the
transition temperature of the interacting gas remains a controversial
subject, even in the uniform case. In this note, we present a calculation
that is simple and transparent, in the hope that it will settle the
controversy.

In the ideal Bose gas, the condition determining the fugacity $z$ in the gas
phase is 
\begin{equation}
\lambda ^{3}n=g_{3/2}\left( z\right)  \label{fugacity}
\end{equation}
where $n$ is the particle density, $\beta =(k_{B}T)^{-1}$,
$\lambda =\sqrt{2\pi \beta\hbar^{2}/m}$ is the thermal wavelength, and 
\begin{equation}
g_{\alpha }(z)\equiv \sum_{\ell =1}^{\infty }\frac{z^{\ell }}{\ell ^{\alpha }}\ .
\end{equation}
No single-particle state has a macroscopic occupation in this phase.
However, the function $g_{3/2}\left( z\right)$ is monotonic, and bounded
by $ g_{3/2}(1)=\zeta (3/2)=2.612$. Thus, particles must go into the
zero-momentum state when $\lambda ^{3}n>\zeta (3/2)$, making this state
macroscopically occupied. This gives the transition temperature of the ideal
Bose gas at a fixed density $n$: 
\begin{equation}
T_{0}=\frac{2\pi \hbar ^{2}}{mk_{B}}\left[ \zeta (3/2)\right] ^{-2/3}\ .
\end{equation}

Now consider a uniform imperfect Bose gas with repulsive interactions with
equivalent hard-sphere diameter $a$ (S-wave scattering length). We denote
the transition temperature by $T_{c}$, and its fractional shift by 
\begin{equation}
\frac{\Delta T}{T_{0}}\equiv \frac{T_{c}-T_{0}}{T_{0}}\ .
\end{equation}
In an early statement on the subject \cite{Huang_Studies}, it was argued
that $\Delta T>0$, since a spatial repulsion leads to momentum-space
attraction\cite{Huang_book}, and this would make the imperfect gas more
ready to condense. However, a Hartree-Fock calculation by Fetter and Walecka 
\cite{Fetter}, and one by Girardeau \cite{Girardeau} based on a mean-field
method, yield the opposite sign $\Delta T<0$. A calculation of the grand
partition to one-loop order by Toyoda \cite{Toyoda} also yields a negative
sign. Specifically, Toyoda obtains 
\begin{equation}
\Bigl( \frac{\Delta T}{T_{0}}\Bigr) _{\text{Toyoda}}=-K_{0}(na^{3})^{1/6}
\end{equation}
where 
\begin{equation}
K_{0}={ 8\sqrt{2\pi} \over 3\left[ \zeta (3/2)\right] ^{2/3} }=3.527\ .
\label{number}
\end{equation}
Barring calculational errors, this must be considered reliable, since it is
the lowest-order result of a systematic expansion. All subsequent
calculations, alas, yield answers different from this and from one another.
Stoof \cite{Stoof} gives $\Delta T\sim a^{3/2}$, while Bijlsma and Stoof 
\cite{Bijlsma} obtain $\Delta T\sim a^{1/2}$. A numerical calculation based
on Monte-Carlo simulation by Gr\"{u}ter et~al.\negthinspace\cite{Gruter} gives $\Delta
T/T_{0}=c_{0}(na^{3})^{\gamma }$, where $c_{0}=0.34\pm 0.06$, and $\gamma
=0.34\pm 0.03$. A recent calculation involving some mean-field assumptions 
\cite{Holzmann} gives $\Delta T/T_{0}=0.7(na^{3})^{1/3}$. Thus, there is no
consensus on how $\Delta T$ should depend on the scattering length, nor even
the sign!

We shall calculate $T_{c}$ using an extension of (\ref{fugacity}) obtained
some time ago\cite{Huang} via the virial expansion. To lowest order in a
power series in $a$, the result is 
\begin{equation}
\lambda ^{3}n=g_{3/2}\left( z\right) \Bigl[ 1-\frac{4a}{\lambda }
g_{1/2}\left( z\right) \Bigr]\ .  \label{fugacity1}
\end{equation}
As a function of $z$, the right side rises through a maximum at some value $
z=z_{c}$, and then approaches $-\infty $ as $z\rightarrow 1$. Thus, as in
the ideal case, the right side is bounded from above, and the bound occurs
at $z_{c}$.  Since the treatment is valid only when $a/\lambda \ll 1$, we put 
\begin{equation}
z_{c}=1-\delta \qquad (\delta \ll 1)\ .
\end{equation}
The maximum can be located with the help of the expansions \cite{Robinson} 
\begin{eqnarray}
g_{3/2}\left( z_{c}\right) &\approx &\zeta (3/2)-2\sqrt{\pi }\delta ^{1/2} 
\nonumber \\
g_{1/2}\left( z_{c}\right) &\approx &\sqrt{\pi }\delta ^{-1/2}\ .
\end{eqnarray}
We then find 
\begin{equation}
\delta =\frac{2a}{\lambda }\zeta (3/2)\ .
\end{equation}
The critical temperature $T_{c}$ can be found by substituting this value
into (\ref{fugacity1}), with the result 
\begin{equation}
\frac{\Delta T}{T_{0}}=K_{0}(na^{3})^{1/6}
\end{equation}
where $K_{0}$ is \ given in (\ref{number}). Thus, we agree exactly with
Toyoda \ except for the sign. Since the derivation here is simple and
transparent, we believe Toyoda made a sign error.

In order to obtain a shift in the transition
temperature, it is necessary to incorporate interactions
in the gas phase, and that is why the virial expansion is useful.
On the other hand, the quasiparticle picture introduced
via the Bogoliubov transformation is not helpful here, because the
quasiparticles reduce to non-interacting particles at the transition point.

\end{document}